\documentclass[amsmath,notitlepage,amssymb,aps,showkeys,floatfix,prd,a4paper,twocolumn,nofootinbib]{revtex4-2}

\usepackage{graphicx}
\usepackage{amsmath}
\usepackage{epstopdf}
\usepackage{amsfonts,amssymb}
\usepackage[utf8]{inputenc}
\usepackage{pstricks}
\usepackage{color}
\usepackage{multirow}
\usepackage{booktabs}
\usepackage{lipsum}

\usepackage[compat=1.0.0]{tikz-feynman}
\usepackage{simplewick}
\usepackage{color, colortbl}
\definecolor{Gray}{gray}{0.9}
\usepackage{float}
\usepackage{tikz-feynman}
\usepackage{feynmp}
\usepackage{forest}

\usepackage{verbatim}    
\usepackage{dcolumn}
\usepackage{bm}
\usepackage{epsfig}
\usepackage{braket}
\usepackage[normalem]{ulem} 
\usepackage{longtable} 

\usepackage{hyperref}

\newcommand{\be}{\begin{equation}}
\newcommand{\ee}{\end{equation}}
\newcommand{\ben}{\begin{eqnarray}}
\newcommand{\een}{\end{eqnarray}}

\def\MeV{\mbox{ MeV}} 
\def\GeV{\mbox{ GeV}}

\def\MeV{\mbox{ MeV}} 
\def\GeV{\mbox{ GeV}} 
 
\newcommand{\pslash}{\not{\hbox{\kern-2.3pt $p$}}}
\newcommand{\pdslash}{\not{\hbox{\kern-2pt $\partial$}}}

\begin{document}

\title{Interaction of exotic states in a hadronic medium: the $Z_c(3900)$ case}

\author{ L. M. Abreu}
\email{luciano.abreu@ufba.br}
\affiliation{ Instituto de F\'isica, Universidade Federal da Bahia,
Campus Universit\'ario de Ondina, 40170-115, Bahia, Brazil}
\affiliation{Instituto de F\'{\i}sica, Universidade de São Paulo 
, São Paulo, SP, Brazil}

\author{R. O. Magalhães}
\email{rodrigomagalhaes@ufba.br}
\affiliation{ Instituto de F\'isica, Universidade Federal da Bahia,
Campus Universit\'ario de Ondina, 40170-115, Bahia, Brazil}

\author{F. S. Navarra}
\email{navarra@if.usp.br}
\affiliation{Instituto de F\'{\i}sica, Universidade de São Paulo,  
Rua do Mat\~ao 1371,  05508-090
Cidade Universit\'aria, São Paulo, SP, Brasil} 


\author{H. P. L. Vieira}
\email{hildeson.paulo@ufba.br}
\affiliation{ Instituto de F\'isica, Universidade Federal da Bahia,
Campus Universit\'ario de Ondina, 40170-115, Bahia, Brazil}

\begin{abstract}

We investigate the interactions of the charged exotic state $Z_c(3900)$ in a hadronic medium composed of light mesons. We study processes such as $Z_c \pi \to D\bar{D}$, $Z_c \pi \to D^*\bar{D}^*$, $Z_c \pi \to D\bar{D}^*$ and the inverse ones. Using effective Lagrangians and form factors calculated with QCD sum rules (treating the $Z_c(3900)$ as a  tetraquark) we estimate the vacuum and thermally-averaged cross-sections of these reactions. We find that the $Z_c(3900)$ has relatively large interaction cross sections with the constituent particles of the hadronic medium.
After that,  we use the production and suppression cross sections in a rate equation to estimate the time evolution of the $Z_c$ multiplicity. We include the $Z_c$ decay and regeneration terms. The coalescence model is employed to compute the initial $Z_c$ multiplicity for the compact tetraquark configuration. Our results indicate  that the combined effects of hadronic interactions, hydrodynamical expansion, decay and regeneration affect the final yield, which is bigger than the initial value. Besides, the dependence of the $Z_c$ final yield with the centrality, center-of-mass energy and the charged hadron multiplicity measured at midrapidity $[dN_{ch}/d\eta \,(\eta<0.5)]$ is also investigated.


\end{abstract}

\maketitle

\section{Introduction}
\label{Introduction} 

Hadron physics has definitively entered a new era since the early 2000's, with the discovery of new states that do not fit in the conventional definition of hadrons as  $q \bar{q}$ mesons and  $q q q $ baryons. For a recent and comprehensive review  we refer the reader to Refs.~\cite{brambilla2019xyz,wang2022radiative}. Here we will focus on the $Z_c(3900)$ state. The charged components $Z_c(3900)^{\pm}$ were observed in 2013 by the BESIII collaboration as a peak in the invariant mass distribution of the system $ \pi^{\pm} J/\psi $ in the $e^+ e^- \to \pi^{+} \pi^+ J/\psi $ reaction at the center-of mass energy of $\sqrt{s} = 4.16 \GeV$~\cite{ablikim2013observation}. The Belle 
collaboration has seen these states in  $e^+ e^-$ collisions at $\sqrt{s}$ in the range $9.46 - 10.86 \GeV $~\cite{liu2013study}. Subsequent partial wave analysis has established their quantum numbers as $J^{P} = 1^{+}$ with statistical significance of more than $7.0 \sigma$~\cite{BESIII:2017bua}. 
In addition, the neutral component $Z_c(3900)^{0}$  was seen in the $ \pi^{0} J/\psi $ invariant mass of the $e^+ e^- \to \pi^{0} \pi^0 J/\psi $ reaction at $\sqrt{s} = 4.23, 4.26$ and  $4.36 \GeV$ by BESIII ~\cite{ablikim2015neutro} (see also \cite{ablikim2015observation,ablikim2014observation}); and at $4.17 \GeV$ by using the CLEO-c detector~\cite{xiao2013observation}. 
According to the Review of Particle Physics (RPP) 2022~\cite{Workman:2022ynf}, the $Z_c(3900)$ averaged mass and width are respectively
\begin{align}
    M_{Z_c} =3887.1\pm 2.6~\MeV,
    \Gamma_{Z_c} = 28.4 \pm 2.6 \MeV. 
\label{masswidth}
\end{align} 
These properties make the $Z_c(3900)$ incompatible with a mesonic structure and with the usual quark-model predictions, giving it the status of the first charged tetraquark state.


On the theoretical side there is a lively debate on the intrinsic nature of the $Z_c(3900)$, its production mechanisms and decay modes. Concerning its internal structure, several possibilities have been considered such as the hadronic molecule  (see i.e. Refs.~\cite{PhysRevD.88.054007, chen2016hidden, ChenAndDong, sun2011z, sun2012note, wang2014possible, chen2015mass, wilbring2013electromagnetic, dong2013strong, dong2014selected, li2014more, gutsche2014radiative, esposito2015probing, ke2013z,Ji:2022uie,Yan:2023bwt}); the compact tetraquark  \cite{maiani2005diquark, maiani2013j, Ali:2011ug, deng2014interpreting, Wang2018, dias2013z}; kinematical singularities  such as threshold cusp effects~\cite{Chen:2013coa,Swanson:2014tra} or triangle singularities~\cite{Liu:2015taa,Szczepaniak:2015eza,Llanes-Estrada:2022qC}, and so on.  In particular, in  Ref.~\cite{Ji:2022uie} the $Z_c(3900)$ was considered as a state belonging to the multiplet of $S-$wave hadronic molecules formed by a pair of ground state charmed and anticharmed mesons. 
Very recently, in Ref.  \cite{Yan:2023bwt} the $Z_c(3900)$  was treated as a $ \pi J/\psi - D\bar D^*$ coupled-channel system in a covariant framework and the invariant masses extracted from the  BESIII data and from lattice simulations were fitted. The probability of the  $D\bar D^*$ component (calculated via compositeness condition) was found to be less than 0.5, suggesting that other hadronic components or compact quark state cores could be also important in the structure  of the $Z_c(3900)$.


In this work we will present predictions of some observables which might be useful for the analysis of the $Z_c(3900)$ properties and for the  determination of its internal structure. We will continue to develop our program on exotics in 
heavy-ion collision (HIC), which has already delivered interesting and potentially useful results. In these collisions, we expect to observe  an enhancement in the number of exotic states, since a large number of heavy quarks are produced in the initial stages. After the collision and the formation of the quark-gluon plasma (QGP) phase, the system expands and cools down, passing by a quark-hadron mixed phase and then reaching the hadronization point where a hot hadron gas is produced. During this transition the heavy quarks coalesce into bound states, mostly mesons and baryons, but also into multiquark states.  Indeed,  this expectation was recently confirmed by the observation of the most famous exotic state, the $X(3872)$, in Pb-Pb collisions at
$\sqrt{s_{NN}} = 5.02$ TeV by the CMS Collaboration~\cite{CMS:2021znk}. The obtained prompt $X(3872)$ production rate is about 10 times larger than that observed in p-p collisions. This motivates us to continue to regard HICs as a promising experimental environment to study the nature of exotic hadrons. 

The final yield of most of the conventional hadrons in HICs has been estimated with the hadron statistical and coalescence models~\cite{EXHIC} which do not include the interactions during the hadron gas phase. However, as previous works suggest~\cite{Abreu:2022lfy, abreu2018impact, abreu2022exotic}, hadronic interactions might strongly affect the final yield.  Any hadronic state can be suppressed due to its interaction with other hadrons of the medium, or be formed from the interaction of other particles. This issue becomes more crucial for the exotic states, because different structures (hadron molecule or compact tetraquark) have distinct spatial configurations, which lead to different interactions and cross sections. 
Thus, based on preceding analyses which tell us that these interactions are strongly dependent on the configuration, in the present work we investigate the interactions of the $Z_c(3900)$ with the  hadronic medium formed in heavy-ion collisions.
In particular, we analyze the $Z_c(3900)$  suppression and production through processes such as $Z_c \pi \to D\bar{D}$, $Z_c \pi \to D^*\bar{D}^*$, $Z_c \pi \to D\bar{D}^*$, and the respective inverse reactions. By making use of effective Lagrangians and form factors calculated with  QCD sum rules, we estimate the vacuum and thermally-averaged cross-sections for these reactions. 
After that, these cross sections are employed in a kinetic equation in order to study the time evolution of the $Z_c$ multiplicity. The small $Z_c$ lifetime is also considered by including the $Z_c$ decay and regeneration terms through an effective coupling, determined from the experimental data. The coalescence model is employed to compute the initial $Z_c$ multiplicity for the compact tetraquark configuration. Besides, the dependence of the $Z_c$ final yield on the centrality, center-of-mass energy and the charged hadron multiplicity measured at midrapidity $[dN_{ch}/d\eta \,(\eta<0.5)]$ is also predicted.

In the next section we will briefly describe
the effective Lagrangian formalism and estimate the vacuum and thermally-averaged cross sections. In section III we present our approach for the evaluation of the $Z_c$ multiplicity and discuss the results obtained. Finally, section IV is devoted to the concluding remarks.


\section{Interactions of $Z_c(3900)$ with pions}
\label{amplitude}

\subsection{Effective Lagrangians and reactions}

\begin{figure} [!ht]

\begin{align*}
\begin{tikzpicture}
\begin{feynman}
\vertex (a1) {$Z_c (p_1)$};
	\vertex[right=1.5cm of a1] (a2);
	\vertex[right=1.cm of a2] (a3) {$\bar{D} (p_3)$};
	\vertex[right=1.4cm of a3] (a4) {$Z_c (p_1)$};
	\vertex[right=1.5cm of a4] (a5);
	\vertex[right=1.cm of a5] (a6) {$\bar{D} (p_{3})$};
\vertex[below=1.5cm of a1] (c1) {$\pi (p_2)$};
\vertex[below=1.5cm of a2] (c2);
\vertex[below=1.5cm of a3] (c3) {$D (p_4)$};
\vertex[below=1.5cm of a4] (c4) {$\pi (p_2)$};
\vertex[below=1.5cm of a5] (c5);
\vertex[below=1.5cm of a6] (c6) {$D (p_4)$};
	\vertex[below=2cm of a2] (d2) {(a)};
	\vertex[below=2cm of a5] (d5) {(b)};
\diagram* {
(a1) -- (a2), (a2) -- (a3), (c1) -- (c2), (c2) -- (c3), (a2) -- [fermion, edge label'= $D^{*}$] (c2), (a4) -- (a5), (a5) -- (c6), (c4) -- (c5), (c5) -- (a6), (a5) -- [fermion, edge label'= $\bar{D}^{*}$] (c5)
}; 
\end{feynman}
\end{tikzpicture}
\end{align*}

\begin{align*}
\begin{tikzpicture}
\begin{feynman}
\vertex (a1) {$Z_c (p_1)$};
	\vertex[right=1.5cm of a1] (a2);
	\vertex[right=1.cm of a2] (a3) {$\bar{D}^{*} (p_3)$};
	\vertex[right=1.4cm of a3] (a4) {$Z_c (p_1)$};
	\vertex[right=1.5cm of a4] (a5);
	\vertex[right=1.cm of a5] (a6) {$\bar{D}^{*} (p_{3})$};
\vertex[below=1.5cm of a1] (c1) {$\pi (p_2)$};
\vertex[below=1.5cm of a2] (c2);
\vertex[below=1.5cm of a3] (c3) {$D^{*} (p_4)$};
\vertex[below=1.5cm of a4] (c4) {$\pi(p_2)$};
\vertex[below=1.5cm of a5] (c5);
\vertex[below=1.5cm of a6] (c6) {$D^{*} (p_4)$};
	\vertex[below=2cm of a2] (d2) {(c)};
	\vertex[below=2cm of a5] (d5) {(d)};
\diagram* {
(a1) -- (a2), (a2) -- (a3), (c1) -- (c2), (c2) -- (c3), (a2) -- [fermion, edge label'= $D$] (c2), (a4) -- (a5), (a5) -- (c6), (c4) -- (c5), (c5) -- (a6), (a5) -- [fermion, edge label'= $\bar{D}$] (c5)
}; 
\end{feynman}
\end{tikzpicture}
\end{align*}

\begin{align*}
\begin{tikzpicture}
\begin{feynman}
\vertex (a1) {$Z_c (p_1)$};
	\vertex[right=1.5cm of a1] (a2);
	\vertex[right=1.cm of a2] (a3) {$D (p_3)$};
\vertex[below=1.5cm of a1] (c1) {$\pi (p_2)$};
\vertex[below=1.5cm of a2] (c2);
\vertex[below=1.5cm of a3] (c3) {$\bar{D}^{*} (p_4)$};
	\vertex[below=2cm of a2] (d2) {(e)};
\diagram* {
(a1) -- (a2), (a2) -- (a3), (c1) -- (c2), (c2) -- (c3), (a2) -- [fermion, edge label'= $\bar{D}^{*}$] (c2)
}; 
\end{feynman}
\end{tikzpicture}
\end{align*}
\caption{Diagrams contributing to the following processes: $Z_c\pi \to D\bar{D}$ (\textbf{a}) and (\textbf{b}), $Z_c\pi \to D^*\bar{D}^*$ (\textbf{c}) and (\textbf{d}), and $Z_c\pi \to D\bar{D}^*$ (\textbf{e}). The charges of the particles are not specified. The momenta of the particles in the  initial (final) state are denoted as $p_1$ and $p_2$ ($p_3$ and $p_4$).}
\label{fig:diagrams}
\end{figure}
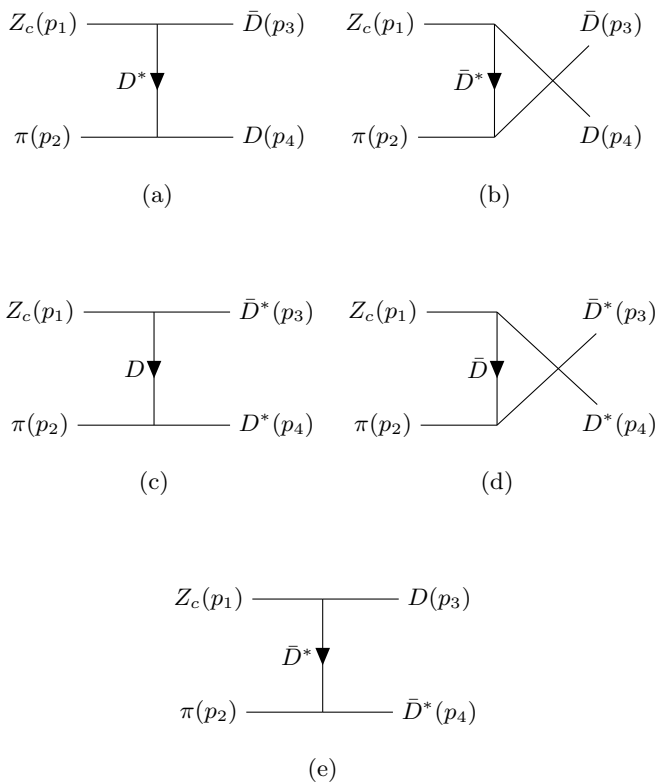

We are interested in the interactions of the $Z_c(3900)$ state (denoted as $Z_c$ from now on) with a hadronic medium constituted of light hadrons. In principle we should include the interactions with all light mesons, such as  $\pi$, $K$, $\eta$, $\rho$, $\omega$, ...etc. However, from our previous experience with other multiquark states (such as the $\chi_{c1}(3872)$, $\chi_{c1}(4274)$, 
$Z_{cs}(3895)$, $T^+_{cc}$...) we learned that the pions are the most important particles and are enough for a first description of 
the hadronic gas. Therefore here we consider only the reactions involving the $Z_c$ and pions. Specifically, for the case of $Z_c$ 
suppression induced by a pion, we will look at the reactions $Z_c \pi \to D\bar{D}$, $Z_c \pi \to D^*\bar{D}^*$ and $Z_c \pi \to D\bar{D}^*$, taking into account the lowest-order Born contributions. In Fig.~\ref{fig:diagrams} the diagrams of these reactions are depicted. The corresponding transition amplitudes will be determined by means of the following effective Lagrangians~\cite{ChenAndDong, dong2013strong, gutsche2014radiative, abreu2022interactions, cho2013hadronic, Nielson2007, bracco2012charm, oh2001j, carvalho2005hadronic}:
\begin{align}
\label{lagrangianas}
&\mathcal{L}_{Z_c DD^*} = g_{Z_cDD^*}D^*_\mu Z_c^\mu \cdot \vec{\tau}\bar{D} + H.c, \nonumber \\
	&\mathcal{L}_{\pi DD^*} = ig_{\pi D D^*} D^{*}_\mu \vec{\tau} \cdot (\bar{D} \partial^\mu \vec{\pi} - \partial^{\mu} \bar{D} \vec{\pi}) + H.c.,  \nonumber \\
	&\mathcal{L}_{\pi D^*D^*}=-g_{\pi D^*D^*}\varepsilon^{\mu\nu\alpha\beta}\partial_\mu D^*_\nu \pi \partial_\alpha\bar{D}^*_\beta,
\end{align}
where $g_{Z_cDD^*}$, $g_{\pi D D^*}$ and $g_{\pi D^*D^*}$ are coupling constants; $\vec{\pi}$ and $\vec{Z}_c^\mu$ represent the pion and $Z_c$ isospin triplets; $\vec{\tau}$ denote the Pauli matrices in the isospin space; $D^{(*)} $ and $\bar D^{(*)} $ are the isospin doublets for the pseudoscalar (vector) charmed mesons. 


The transition amplitudes associated to the reactions shown in Fig.~\ref{fig:diagrams} can be determined from the effetive Lagrangians in Eq.~(\ref{lagrangianas}) and are given by
\begin{align} \label{eq: amplitude sum}
&\mathcal{M}(Z_c \pi \to D\bar{D}) = \mathcal{M}^{(a)}+\mathcal{M}^{(b)}, \nonumber \\
&\mathcal{M}(Z_c \pi \to D^*\bar{D}^*) = \mathcal{M}^{(c)}+\mathcal{M}^{(d)}, \nonumber \\
&\mathcal{M}(Z_c \pi \to D\bar{D}^*) = \mathcal{M}^{(e)},
\end{align}
where ${M}^{(i)}$ represents the amplitude coming from the specific process $(i)=(a),..,(e)$; these expressions are explicitly written as 
\begin{eqnarray}
\mathcal{M}^{(a)}& = & g_{Z_cDD^*}g_{\pi D D^*} \epsilon_1^{\mu} (p_2+p_4)^\nu \nonumber \\ & &  
\times \left( g_{\mu \nu}-\dfrac{(p_1-p_3)_{\mu} (p_1-p_3)_{\nu}}{m_{D^*}^2}\right)\frac{1}{t-m_{D^*}^2} , \nonumber \\ 
\mathcal{M}^{(b)} & = & - g_{Z_cDD^*}g_{\pi D D^*}\epsilon_1^{\mu} \times(p_2+p_3)^\nu \nonumber \\& &
 \times \left( g_{\mu \nu}-\dfrac{(p_1-p_4)_{\mu} (p_1-p_4)_{\nu}}{m_{\bar{D}^*}^2}\right) \frac{1}{u-m_{\bar{D}^*}^2} , \nonumber \\ 
\mathcal{M}^{(c)} & = & g_{Z_cDD^*} g_{\pi DD^*}\epsilon_1^\mu \epsilon^{*}_{3\mu}\epsilon^{*}_{4\nu} 
\left( \frac{1}{t-m_{D}^2} \right)(2p_2-p_4)^\nu, \nonumber \\
\mathcal{M}^{(d)} & = & -g_{Z_cDD^*}g_{\pi DD^*}\epsilon_1^\mu \epsilon^{*}_{4\mu} \epsilon^{*}_{3\nu} 
\left( \frac{1}{u-m_{\bar{D}}^2} \right)(2p_2-p_3)^\nu, \nonumber \\
\mathcal{M}^{(e)} & = & -g_{Z_cDD^*} g_{\pi D^*D^*} \varepsilon^{\mu\nu\alpha\beta}\epsilon_1^\gamma\epsilon^{*}_{4\nu} \nonumber \\ & &
\times \left( g_{\gamma \beta}-\dfrac{(p_1-p_3)_{\gamma} (p_1-p_3)_{\beta}}{m_{\bar{D}^{*}}^2} \right) \nonumber \\ & & 
\times \frac{1}{t-m_{\bar{D}^{*}}^2}p_{4\mu} (p_4-p_2)_{\alpha},
\end{eqnarray}
where $s, u$ and $t$ are the Mandelstam variables, defined as $ s=(p_1+p_2)^2, t=(p_1-p_3)^2 $ and $ u=(p_1-p_4)^2$; and $\epsilon_i^{(\ast)} \equiv \epsilon^{(\ast)} (p_i)$ is the polarization vector associated to the corresponding vector meson.  


\subsection{Form factors and coupling constants}
\label{formfactor}

In effective Lagrangian approaches we often use form factors to incorporate features associated to the finite size of the hadrons and also to 
prevent the artificial growth of the cross sections with the energy. A form factor can be obtained from the three-point function constructed 
with three meson currents. In QCD sum rules (QCDSR) this three-point function is evaluated first writing the currents in terms of the quark fields, contracting them and performing an Operator Product Expansion (OPE) of the resulting propagators. Then we evaluate the same three-point function using hadronic degrees of freedom. The sum rule is the equation obtained equating the two expressions of the same three-point 
function. In this equation the unknown is the form factor. More precisely, in a generic vertex of three mesons $M_1$, $M_2$ and $M_3$ the 
form factor is the function $g_{M_1 \,  M_2 \, M_3} (p,p')$ ($p$ and $p'$ being the external 4-momenta) and it is evaluated in terms
of the QCD parameters (quark masses and couplings) as well as of the meson masses and decay constants (see a detailed discussion in Ref.~\cite{abreu2022interactions}).  Here we benefit from previous works in which all the form factors required for the present calculation have been computed with QCDSR.  

In the case of the vertices $D\pi D^*$ and $ D^* \pi D^*$, the corresponding form factors have been calculated with QCDSR in Ref.~\cite{bracco2012charm}; they were parametrized as: 
\begin{equation}
(I) \; g_{M_1M_2M_3}(Q^2)=\frac{A}{Q^2+B}
\label{FI}
\end{equation}
and
\begin{equation}
(II) \; g_{M_1M_2M_3(Q^2)}=Ae^{-(Q^2/B)},
\label{FII}
\end{equation}
where $M_1$ is the off-shell meson in the vertex and $Q^2=-q^2$ its Euclidean four momentum; the parameters $A$ and $B$ are given in the Table \ref{tab: form factors}.

\begin{table}[H]
\centering
\begin{tabular}{|c|c|c|c|}
\hline 
$M_1 M_2 M_3$ & Form & $A$ & $B$ \\ 
\hline 
$D \pi D^*$ & I & 126 & 11.9 \\
\hline
$D^* \pi D^*$ & II & 4.8 & 6.8 \\
\hline
\end{tabular} 
\caption{Parameters of the form factors in the vertices with the mesons $M_1$, $M_2$ and  $M_3$ \cite{abreu2022interactions, bracco2012charm}.}
		\label{tab: form factors}
\end{table}

In Ref.~\cite{dias2013z} the $Z_c^+  D^+ \bar{D}^{*0} $ and $Z_c^+  \bar{D}^0 D^{*+} $ vertices were studied with  QCDSR. The 
 $Z_c$ was treated as a genuine tetraquark  $(\bar{c}\bar{q})(cq)$ with  non-trivial color structure. The resulting form factor 
 (valid for both vertices) was found to be
\begin{equation} \label{form factor Zc}
g_{Z_cDD^*}(Q^2)=g_{Z_cDD^*}e^{-g(Q^2+m_D^2)}
\end{equation}
with $g=0.076 \, \text{GeV}^{-2}$.
Hence, from Eqs.~(\ref{FI}), (\ref{FII}) and (\ref{form factor Zc}) the coupling constants can be fixed with $g_{M_1M_2M_3}(-M_1^2)$; they are: 
\begin{align} \label{Zc coupling constants}
g_{\pi DD^*} &= 14.0 \pm 1.5 \, \nonumber \\
g_{\pi D^*D^*} &=(8.6 \pm 1.0) \, \text{GeV}^{-1}, \nonumber \\
g_{Z_c DD^*} & = (2.5 \pm 0.3)\,\, \text{GeV}. 
\end{align}
The coupling $g_{Z_c DD^*} $ involving the neutral component $Z_c^0$ is not yet known. Notwithstanding, due to the proximity of the masses between the neutral and charged components of the $Z_c$, we will assume that the coupling $g_{Z_c DD^*} $ is also valid for the neutral component.  


\subsection{Cross sections}
\label{CrossSection}

The isospin-spin-averaged cross section in the center of mass (CM) 
frame for a given reaction $ab\rightarrow cd$ in Eq.~(\ref{eq: amplitude sum}) is given by:
\begin{equation} \label{seção de choque total}
\sigma_{ab \to cd} = \frac{1}{64 g_a g_b \pi^2 s} \frac{|\vec{p}_{cd}|}{|\vec{p}_{ab}|} \int d\Omega \overline{\sum} |\mathcal{M}_{ab \to cd} |^2,
\end{equation} 
where $g_{a,b}=(2I_{a,b}+1)(2S_{a,b}+1)$ is the degeneracy factor of the 
particles in the initial state; $s$ is the squared CM energy;         
$| \vec{p}_{ab} |$ and $| \vec{p}_{cd} |$ are the absolute values of the three-momenta  
in the CM frame of the initial and final particles; $\overline{\sum}$ denotes the summation
over the spin and isospin of the initial and final states, which can be rewritten in the particle basis as
\begin{equation}
 \displaystyle  \sum_{I} |M_{ab\rightarrow cd}|^2\rightarrow\sum_{Q_c,Q_d} 
|M_{ab\rightarrow cd}^{(Q_c,Q_d)}|^2.
\end{equation}
with $Q_c,Q_d$ being the explicit charges of the particles in the final state.
For the inverse processes in which the $Z_c(3900)$ might be produced, i.e. $D\bar{D} \to Z_c\pi$, $D^*\bar{D}^* \to Z_c\pi$ and 
$D\bar{D}^* \to Z_c\pi$, the cross sections are calculated with the help of the detailed balance equation, 
\begin{equation}
g_ag_b|\vec{p}_{ab}|^2\sigma_{ab \to cd}(s) = g_cg_d|\vec{p}_{cd}|^2\sigma_{cd \to ab}(s).
\label{detbal}
\end{equation}





Now we can present and discuss the results. We remark that the estimates have been done with the isospin-averaged masses reported in~\cite{Workman:2022ynf}. Since the coupling constants are the same (see discussion in section \ref{formfactor}),  we assume that the 
charged $Z_c^{\pm}$ and neutral $Z_c^0$ components have identical contributions. So, the charges will not be explicit. Also, the 
results are presented in terms of bands associated to the smallest and largest possible values of the coupling constants.

\begin{figure}[h!]
\centering
\includegraphics[{width=0.9\linewidth}]{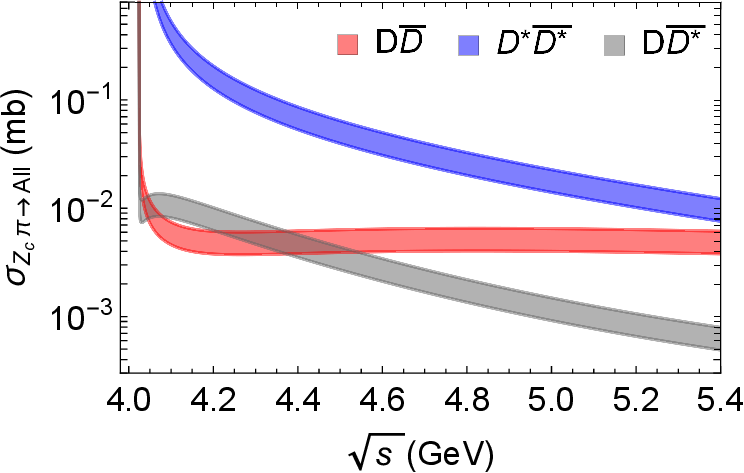}\\ 
\vskip0.5cm
\includegraphics[{width=0.9\linewidth}]{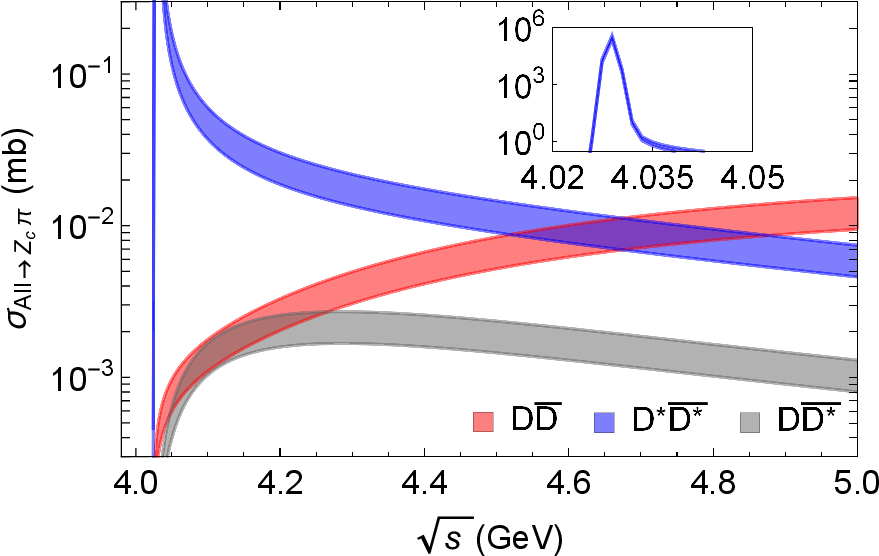}
\caption{Top: cross sections for the suppression processes $Z_c \pi \to D^{(*)}\bar{D}^{(*)}$ as functions of $\sqrt{s}$. Bottom: cross sections for the corresponding inverse reactions.}
    \label{Fig: CrossSection}
\end{figure}

In Fig.~\ref{Fig: CrossSection} we show the suppression processes $Z_c \pi \to D^{(*)}\bar{D}^{(*)}$ as functions of $\sqrt{s}$ and the corresponding inverse reactions.  All the absorption processes are exothermic. The cross sections are very large close to the threshold and decrease with increasing  energy  $\sqrt{s}$. At moderate values of CM energy the reaction with final state $ D^*\bar{D}^*$ yields the most 
relevant contribution, around one order of magnitude larger than the others. This might be partially understood from the dynamics, as the exchanged charmed meson in this reaction is lighter than the one in  other processes. 

\begin{figure}[h!]
\centering
\includegraphics[{width=1\linewidth}]{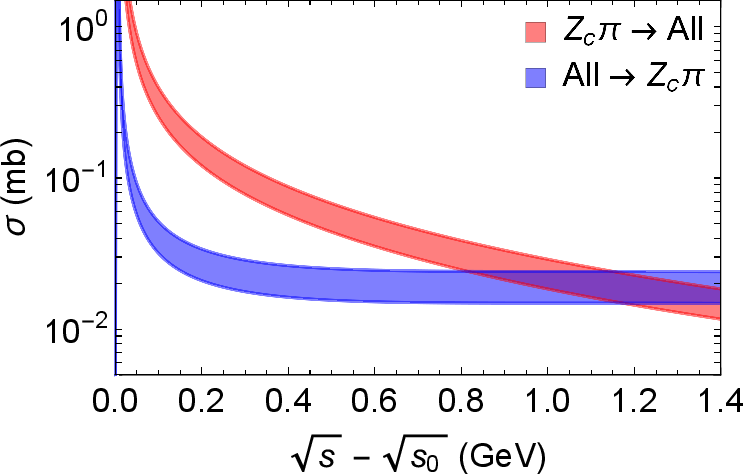}
\caption{Sum of all cross-sections for the $Z_c$ absorption and production processes as functions of $\sqrt{s}-\sqrt{s_0}$, 
where $\sqrt{s_0}$ is the corresponding threshold of each channel. $All $ represents the sum 
$\sigma(Z_c \pi \to D\bar{D})+ \sigma(Z_c \pi \to D^*\bar{D}^*)+\sigma(Z_c \pi \to D\bar {D}^*)$ for the case of absorption processes 
and an equivalent  expression for the inverse ones. }
\label{fig: CrossSectionSum}
\end{figure}

As we can see from Fig.~\ref{Fig: CrossSection}, the $Z_c$ production processes  are endothermic and the cross sections tend to vanish near the threshold. Then they grow  with $\sqrt{s}$ up to moderate energies.   The channel with $ D^*\bar{D}^*$ in the initial state presents a cusp close to the threshold, due to the small energy difference between the initial and final states.  Comparing the  $Z_c$ absorption and production processes in the region of energies relevant for heavy-ion collisions ($\sqrt{s} -\sqrt{s_0} <  0.6$ GeV),
the absorption cross sections are greater than the production ones.  The difference comes from the phase space and the degeneracy factors encoded in Eq.~(\ref{detbal}). This feature becomes more evident in Fig.~\ref{fig: CrossSectionSum}, where the sum of all cross-sections for the $Z_c$  absorption and production processes are shown as a function of $\sqrt{s}-\sqrt{s_0}$ ($\sqrt{s_0}$ is the corresponding threshold of each channel). 

\subsection{Thermally-averaged cross sections}
\label{ThermalAverage}

In view of our interest in a heavy-ion collision environment, in which the medium effects become relevant and the $Z_c$ may interact with light hadrons, it is necessary to evaluate the thermally-averaged cross sections, as the collision energy is related to the medium temperature. 
The cross-section averaged over the thermal distributions of the 
particles participating in the reactions can be defined as the convolution of vacuum cross-sections and the momentum  distributions. Strictly speaking, for a given reaction of the type $ab \to cd$, it is given by~\cite{cho2013hadronic, ABREU2016303,Abreu:2022lfy, abreu2018impact, abreu2022exotic, EXHIC}:
\begin{align}\label{medias termicas}
\langle \sigma_{ab \to cd}v_{ab} \rangle =& \frac{\int d^3\textbf{p}_a  d^3\textbf{p}_b f_a(\textbf{p}_a) f_b (\textbf{p}_b)\sigma_{ab \to cd}v_{ab}}{\int d^3\textbf{p}_a  d^3\textbf{p}_b f_a(\textbf{p}_a) f_b (\textbf{p}_b)}\nonumber \\
=&\frac{1}{4\alpha_a^2K_2(\alpha_a)\alpha_b^2K_2(\alpha_b)}\int_{z_{0}}^{\infty}dzK_1(z) \times \nonumber \\
&\times \sigma(s=z^2T^2)[z^2-(\alpha_a+\alpha_b)^2] \nonumber \\&\times[z^2-(\alpha_a-\alpha_b)^2],
\end{align}
where $v_{ab}$ represents the initial relative velocity of the two interacting particles $a$ and $b$; $f_i(\textbf{p}_i)$ is the momentum distribution (here assumed to be the Bose-Einstein function); $\alpha_i=m_i/T$, where $T$ is the temperature; $ z_0=max(\alpha_a+\alpha_b, \alpha_c+\alpha_d)$;  and $K_1$ and $K_2$ are the modified Bessel functions.

\begin{figure}[h!]
\centering
\includegraphics[{width=0.9\linewidth}]{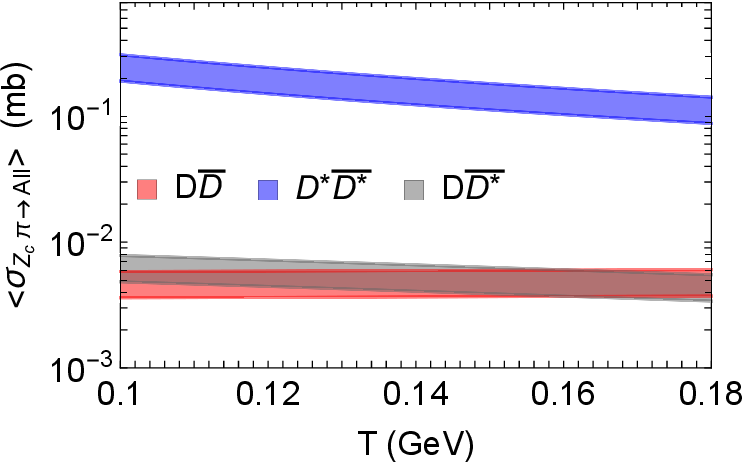}\\
\vskip0.5cm
\includegraphics[{width=0.9\linewidth}]{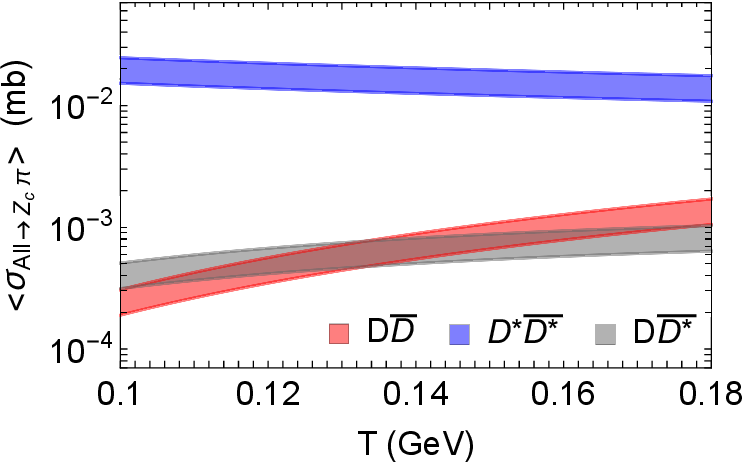}
\caption{Top: Thermally-averaged cross-sections of the processes $Z_c \pi \to D^{(*)}\bar{D}^{(*)}$ as functions of the temperature. 
Bottom:  thermally-averaged cross-sections of the corresponding inverses reactions.}
\label{fig:AvCrsecZcPiAll(Inv)}
\end{figure}

In Fig.~\ref{fig:AvCrsecZcPiAll(Inv)} we plot the thermally-averaged cross-sections for the $Z_c$ absorption and production processes as functions of the temperature. In the case of absorption, all the channels have a weak dependence with the temperature and the one with  $D^*\bar{D}^*$  in the final state dominates. For the production processes, the temperature dependence is more prominent for the $D\bar{D}^*$ 
and $D \bar{D}$ channels, but the $D^*\bar{D}^*$ remains the most important contribution by at least one order of magnitude.

Here is one of the main conclusions of this study: as it happens to most of the other exotic states, the 
$Z_c$ absorption processes are more important than the production reactions. This difference suggests that the hadronic medium may strongly  
affect the final number of produced $Z_c$s. In order to have a concrete prediction of this effect, the above cross sections will be used as 
input in the kinetic equation, whose solution  yields  the final  $Z_c$  multiplicity.


\section{The $Z_c (3900)$ multiplicity}
\label{Abundance}

\subsection{Time Evolution}

Now we  move to the study of the time evolution of the $Z_c$ multiplicity during the hadronic stage of heavy ion collisions. The formalism is the 
same used in our previous works, but for the sake of completeness we briefly describe it here. The thermally-averaged cross sections estimated in the previous section will be employed in the momentum-integrated evolution equation given by~\cite{Abreu1,XProd1,XProd2, Abreu:2017cof,Abreu:2018mnc}:
\begin{eqnarray} 
\frac{ d N_{Z_c} (\tau)}{d \tau} & = & \sum_{\substack{c = D, D^* \\ \bar c = \bar{D}, \bar{D}^{*}}} 
\left[ \langle \sigma_{c \bar c \rightarrow Z_c \pi } 
v_{c \bar c } \rangle n_{\bar{c}} (\tau) N_{c}(\tau) \right. 
    \nonumber 
 \\ & &  \left. - \langle \sigma_{ Z_c \pi\rightarrow c \bar c } v_{ Z_c \pi } 
\rangle n_{\pi} (\tau) N_{Z_c}(\tau)  \right] 
    \nonumber
\\ & & + \langle \sigma_{J/\psi \pi \rightarrow Z_c } v_{J/\psi \pi} \rangle n_{\pi} (\tau) N_{J/\psi} (\tau) 
    \nonumber
\\ & & - \langle \Gamma_{Z_c \rightarrow J/\psi \pi} \rangle N_{Z_c} (\tau) ,
\label{rateequation}
\end{eqnarray}
where $N_{Z_c} (\tau)$ denotes the multiplicity of $Z_c$ at proper time $\tau$; $n_c(\tau), n_{\bar c}(\tau)$ and $n_\phi(\tau)$ are the number densities of charm and light mesons, respectively. The hadron gas is assumed to be in thermal equilibrium, and its constituents have their respective densities following the Boltzmann distribution,
\begin{align}
    n_{i} (\tau) \approx  \frac{1}{2\pi^2}\gamma_i g_im_i^2 T(\tau) K_2\left( \frac{m_i}{T(\tau)}\right),\label{statistical}
\end{align}
with $\gamma_i$, $g_i$, and $m_i$ being the fugacity, degeneracy and mass of the particle $i$, respectively, and $T(\tau)$ the time-dependent temperature. The abundance $N_i(\tau)$ is obtained by multiplying $n_i(\tau)$  by the volume $V(\tau)$.

The last two lines of Eq.~\ref{rateequation} describe the $Z_c$ decay and its regeneration from the daughter particles. The  $Z_c$ has a large decay width and a lifetime shorter than that of the hadron gas phase $(\sim 10 fm/c)$.  According to the Review of Particle Physics (RPP) 2022~\cite{Workman:2022ynf} the $ Z_c \rightarrow  D \bar D$ and  $ Z_c \rightarrow  J/\psi \pi$  reactions have been seen. In Ref.~\cite{dias2013z}  it was shown that the $ Z_c \rightarrow J/\psi \pi $ is the most important decay channel. Thus we will include only 
this channel in our calculations. The other channel is also taken into account in the interactions of the $Z_c$ with pions described in the  previous section. Following the method summarized in Refs.~\cite{Cho:2015qca,Abreu:2020ony}, we employ the effective Lagrangian
\begin{eqnarray}
    \mathcal{L}_{Z_c \psi \pi} = g_{Z_c \psi \pi} \partial_{\mu} \psi_{\nu} (\partial^{\mu} \pi Z^{\nu}_{c} - \partial^{\nu} \pi Z^{\mu}_{c} ).
\label{vertex}
\end{eqnarray}
In this expression $g_{Z_c \psi \pi}$ is the coupling constant to be determined from the decay rate, written as 
\begin{align}
\Gamma_{Z_c \rightarrow J/\psi \pi} (\sqrt{s}) = \frac{1}{8 \pi s} |p_{cm} (\sqrt{s}) | \overline{\sum} |\mathcal{M}_{\Gamma}|^2
\label{decayrate1}
\end{align}
where $p_{cm}$ is the three-momentum in the center-of-mass frame, and $\mathcal{M}$ is the tree-level amplitude of the decay rate,
\begin{align}
    \mathcal{M}_{\Gamma} = -g_{Z_c \psi \pi} p_{2\mu} \epsilon^{*}_{2\nu} [ p^{\mu}_{3} \epsilon_1^{\nu}  - p^{\nu}_{3} \varepsilon_{1}^{\mu} ]
\label{ampl}
\end{align}
with the momenta of the states $Z_c$, $J/\psi$ and $\pi$ being given by $p_1$, $p_2$ and $p_3$, respectively. Using the experimental value of the $Z_c$ decay width in the above formulas we determine the value of the coupling constant: $g_{Z_c \psi \pi} = (3.89 \pm 0.56)$ GeV. The $Z_c$  decay width averaged over the thermal distribution is expressed by
\begin{eqnarray}
    \langle \Gamma_{Z_c \rightarrow J/\psi \pi} \rangle = \Gamma_{Z_c \rightarrow J/\psi \pi} (m_{Z_c}) \frac{K_1 (m_{Z_c} / T)}{K_2 (m_{Z_c} / T)} .
\label{avdecay}
\end{eqnarray}
In the case of regeneration process, the cross section is given by the spin-averaged relativistic Breit-Wigner approximation:
\begin{eqnarray}
\sigma_{J/\psi \pi \rightarrow Z_c} = \frac{g_{Z_c}}{g_{J/\psi}g_{\pi}} \frac{4\pi}{p^{2}_{cm}} \frac{s\Gamma^{2}_{Z_c \rightarrow J/\psi \pi}}{ \left( s - m^{2}_{Z_c} \right)^2 + s\Gamma^{2}_{Z_c \rightarrow J/\psi \pi}}      
\label{BW}
\end{eqnarray}
where $g_{Z_c}$, $g_{J/\psi}$ and $g_\pi$ are the degeneracy factors  of the $Z_c$, $J/\psi$ and $\pi$ respectively. This cross section is averaged over the thermal distribution in the same way as done in the previous section.

\subsection{The expansion of the hadron gas }

To model the hadron gas evolution, we employ the boost invariant Bjorken picture with an accelerated transverse expansion~\cite{Abreu1, Abreu:2017cof, Abreu:2018mnc}. Accordingly, the proper-time dependence of the volume and temperature is:
\ben
V(\tau) & = & \pi \left[ R_C + v_C \left(\tau - \tau_C\right) + 
\frac{a_C}{2} \left(\tau - \tau_C\right)^2 \right]^2 \tau c , \nonumber \\
T(\tau) & = & T_C - \left( T_H - T_F \right) \left( \frac{\tau - 
\tau _H }{\tau _F - \tau _H}\right)^{\frac{4}{5}} .
\label{TempVol}
\een
where $R_C, \upsilon_C, a_C$ and  $T_C$ represent the transverse size, transverse velocity, transverse acceleration and temperature at the critical time $\tau_C$, respectively; $T_H (T_F)$ is the temperature at the end of the mixed phase  time $\tau_H (\tau_F)$. These parameters are fixed for a hadronic medium formed in central Pb-Pb collisions at $\sqrt{s_{NN}} = 5.02$ TeV according to Ref.~\cite{EXHIC}, and are given in Table~\ref{tabela1}. 

The multiplicities of the pions and charmed mesons are also shown in Table~\ref{tabela1}, and are used to fit the fugacities in Eq.~(\ref{statistical}). In the case of charm quarks, since they are produced in the early stages of the collision, their total number $(N_c)$ in charmed hadrons is assumed to be constant during the hadron gas phase, yielding a time-dependent charm-quark-fugacity factor $\gamma _c \equiv \gamma _c (\tau)$ in order to keep $ N_c = n_c (\tau) \times V(\tau) = const.$


    \begin{table}[h!]
       \caption{In the first three lines we list the parameters used in Eq.~(\ref{TempVol})  for central Pb-Pb collisions at $\sqrt{s_{NN}} = 5.02$ TeV~\cite{EXHIC}. In the next two lines we list the multiplicities of the mesons and the quark masses used in the model. 
       In the last line we show the frequency  used in the coalescence model.}
\centering \begin{tabular}{ccc}
\hline \hline $v_C(\mathrm{c})$ & $a_C\left(\mathrm{c}^2 / \mathrm{fm}\right)$ & $R_C(\mathrm{fm})$ \\
0.5 & 0.09 & 11 \\
\hline $\tau_C(\mathrm{fm} / \mathrm{c}) $&$ \tau_H(\mathrm{fm} / \mathrm{c}) $&$ \tau_F(\mathrm{fm} / \mathrm{c})$ \\
7.1 & 10.2 & 21.5 \\
\hline $T_C(\mathrm{MeV}) $&$ T_H(\mathrm{MeV}) $&$ T_F(\mathrm{MeV})$ \\
156 & 156 & 115 \\
\hline $N_\pi\left(\tau_H\right) $&$ N_D\left(\tau_H\right) $&$ N_{D^*}\left(\tau_H\right)$ \\
713 & 4.7 & 6.3  \\
\hline
$N_c$ & $m_c \ [\mathrm{MeV}]$ & $m_q \ [\mathrm{MeV}]$   \\   
14 & 1500  & 350   \\  
\hline $\omega_c[\mathrm{MeV}]$ & \\ 220  &  & \\
\hline \hline \label{tabela1}
\end{tabular}
    \end{table}

To determine the initial conditions for the rate equation (\ref{rateequation})  we employ the coalescence model. It has the relevant feature of carrying information concerning the intrinsic structure of the system (such as angular momentum and the type and number of constituent quarks), because the multiplicity of the hadronic state is computed from the convolution of the density matrix of its constituents and its Wigner Function. Accordingly, the yield of $Z_c$ at $\tau_c$ can be written as~\cite{EXHIC}:
\ben
N_{Z_c}^{Coal} & \approx & g_{Z_c} \prod _{j=1} ^{n} \frac{N_j}{g_j} 
\prod  _{i=1} ^{n-1} 
\frac{(4 \pi \sigma_i ^2)^{\frac{3}{2}} }{V (1 + 2 \mu _i T \sigma _i ^2 )} 
\nonumber \\
& & \times
\left[ \frac{4 \mu_i T \sigma_i ^2 }{3 (1 + 2 \mu _i T \sigma _i ^2 ) }
\right]^{l_i}, 
\label{ZcCoal}
\een
where $g_j$ and $N_j$ are the degeneracy and number of the $j$-th constituent of the $Z_c$ and $\sigma_i=(\mu_i\omega)^{-1/2}$. We assume 
that the hadron internal structure is represented by a harmonic oscillator, the parameter $\omega$ is the oscillator frequency  and $\mu$ is 
the reduced mass. The angular momentum $l_i$ is 0 for a  $S$-wave. Here the $Z_c$ is considered as a $S-$wave  compact tetraquark, with the frequency, the quark numbers and masses summarized in Table~\ref{tabela1}. Hence, putting all these parameters in Eq.~(\ref{ZcCoal}), 
the initial $Z_c$  multiplicity  is 
\begin{align}
    N_{Z_c}(\tau_H)=2.1170  \times 10^{-4} . 
    \label{initialconditions}
\end{align}

\subsection{Size of the System, CM energy and centrality}

As discussed in Refs.~\cite{Abreu1}, it is possible to relate the multiplicity  to other relevant observables, such as the charged-particle pseudorapidity density at mid-rapidity, $[dN_{ch}/d\eta \, (|\eta|<0.5)]$. Using the empirical formula \cite{chiara}
\begin{align}
    T_F=T_{F0} \, e^{-b \mathcal{N} }, 
\end{align}
where $T_{F0}=132.5\MeV$, $b=0.02$ and $\mathcal{N} \equiv [dN_{ch}/d\eta \,(|\eta|<0.5)]^{1/3}$.  Assuming that the hadron gas undergoes a Bjorken-type cooling, then the freeze-out time $\tau_F$ and the freeze-out temperature can be related as follows~\cite{Abreu1}:
\begin{align}
    \tau_F=\tau_H\left( \frac{T_H}{T_{F0}}\right)^3 e^{3b\mathcal{N}}.
    \label{temperaturerelation}
\end{align}
As we can see, $\mathcal{N}$ can be interpreted as an indirect measure of the duration of the hadronic phase, implying that a larger system (with a larger mass number $A$) will yield a larger charged-particle pseudorapidity density (bigger $\mathcal{N}$), generating a longer hadron phase  (bigger $\tau_F$). Hence, the use of Eq.~(\ref{temperaturerelation}) in (\ref{rateequation}) can be seen as an indirect estimate of the dependence of $N_{Z_c}$ with the size of the system.
Besides, $\mathcal{N}$ can also be related to the center-of-mass energy $\sqrt{s}$ and to the centrality of the collision. The empirical formulas connecting $\mathcal{N}$ with $\sqrt{s}$ and the centrality (denoted as $x$, in \%) are given by~\cite{Abreu1}:
\begin{align}
    \frac{d{N}_{ch}}{d\eta}|_{|\eta|<0.5}=    -2332.12+491.69\log(220.06+\sqrt{s}),
    \label{energy}
\end{align}
and
\begin{align}
\left.\frac{d{N}_{ch}}{d\eta}\right|_{|\eta|<0.5}&=2142.16-
85.76x+1.89x^2-0.03x^3+
\notag\\
&+3.67\times10^{-5}x^4-2.24\times 10^{-6}x^5+\notag\\
&+5.25\times10^{-9}x^6. 
\label{centrality}
\end{align}

\subsection{Results and discussion}

\begin{figure}[h!]
\centering
\includegraphics[{width=1\linewidth}]{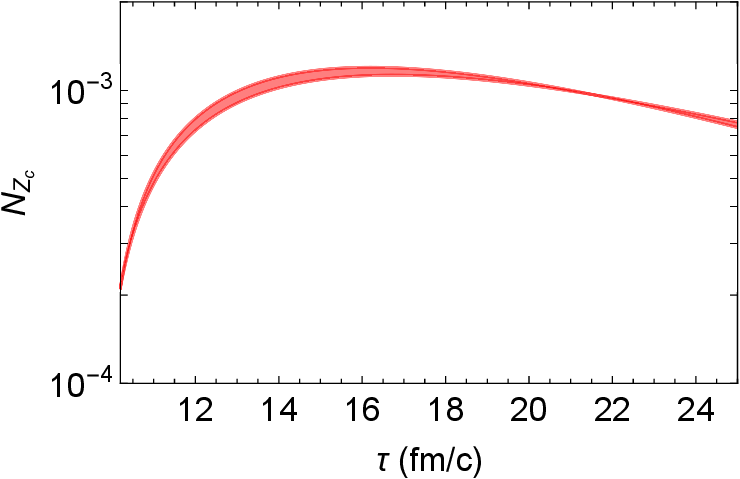}
\caption{Abundance of $Z_c (3900)$ as a function of the proper time in central Pb-Pb collisions at $\sqrt{s_{NN}} = 5.02$ TeV. }
    \label{TimeEvNZ}
\end{figure}

The solution of Eq. (\ref{rateequation}) is presented in Fig.~\ref{TimeEvNZ}. As it can be seen,  at the beginning of the hadron gas phase, the gain terms are dominant and generate an increase of the $Z_c$ multiplicity. However, as the gas expands and cools down the $Z_c$ production rate becomes smaller than the absorption rate, leading to a decrease of the abundance. Notwithstanding, at the freeze-out time the $Z_c$ final yield is about four times larger than the initial value. 


\begin{figure}[h!]
\centering
\includegraphics[{width=1\linewidth}]{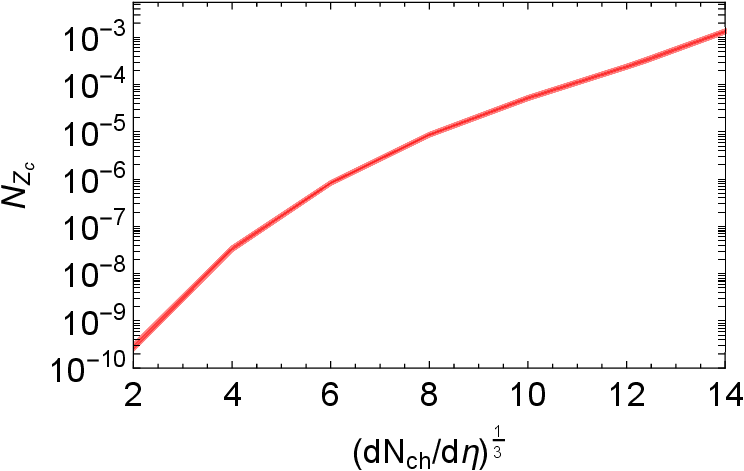}
\caption{$Z_c (3900)$ multiplicity as a function of $\mathcal{N}$.}
    \label{NdNdEta}
\end{figure}

\begin{figure}[h!]
\centering
\includegraphics[{width=1\linewidth}]{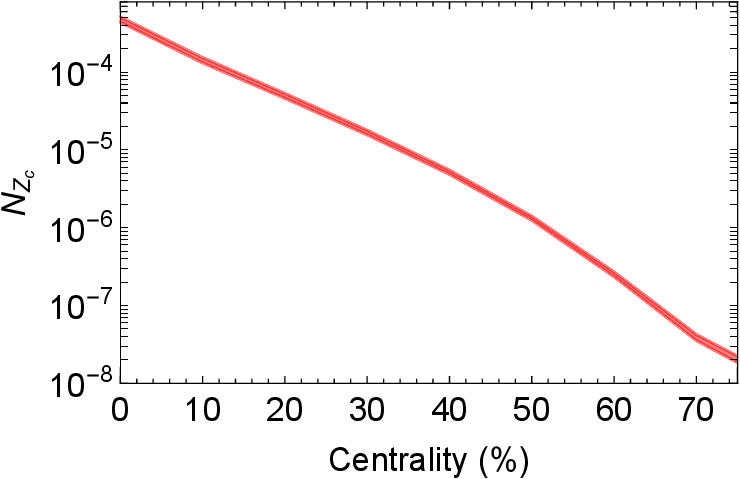}\\
\caption{$Z_c (3900)$ multiplicity as a function of the centrality. }
\label{NCentrality}
\end{figure}

\begin{figure}[h!]
\centering
\includegraphics[{width=1\linewidth}]{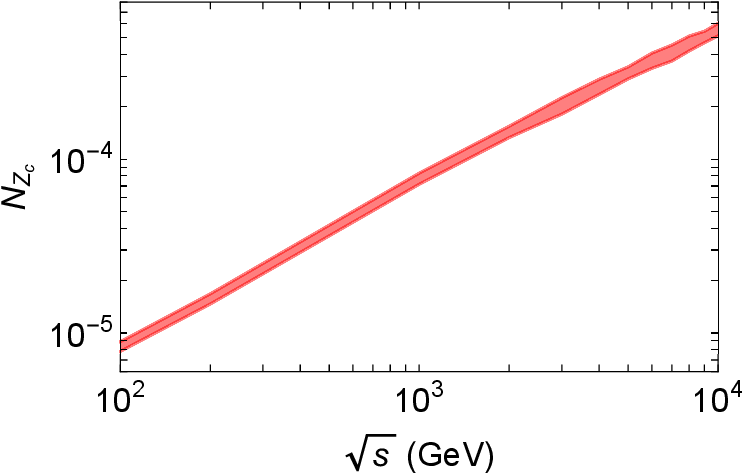}
\caption{$Z_c (3900)$ multiplicity as a function of the CM energy. }
\label{NEnergy}
\end{figure}

In Fig.~\ref{NdNdEta} we show the $Z_c$ multiplicity as a function of $\mathcal{N}$. As expected, it grows with the size of the system.   
Finally, in Fig.~\ref{NCentrality} and ~\ref{NEnergy} we show the $Z_c$ multiplicity as a function of the centrality and energy, obtained by using Eqs.~(\ref{temperaturerelation}), (\ref{energy})  and (\ref{centrality}) in (\ref{rateequation}). We notice that the $Z_c$ final yield decreases from central to peripheral collisions and increases when the system reaches higher energies. 

\section{Concluding remarks}
\label{Conclusions}

Summarizing, the results discussed in the preceding sections suggest that the $Z_c(3900)$ state might suffer sizable interactions with the hadronic medium formed in heavy-ion collisions. 
First, we have studied the vacuum and thermally-averaged cross sections obtained for the $Z_c$ absorption and production processes by comoving pions. They have been obtained from an effective Lagrangian formalism and form factors calculated with QCDSR considering the $Z_c$ as a tetraquark state. 
We have found  that the $Z_c$ absorption processes are much more important than the production reactions. 
With the obtained cross sections  we have studied the time evolution of the $Z_c$ multiplicity, by using the production and suppression thermal cross sections in the rate equation.  In this equation, the $Z_c$ decay and regeneration terms  were included. We have used the coalescence model to compute the initial $Z_c$ multiplicity for the compact tetraquark configuration. Our results indicate  that the combined effects of hadronic interactions, hydrodynamical expansion, decay and regeneration lead to an enhancement of  the $Z_c$ multiplicity at the end of 
the collision. 
As a final comment, it is also worth  mentioning the findings reported in Ref.~\cite{Llanes-Estrada:2022qC}, where the triangle singularity interpretation of the $Z_c$ was discussed in the context of heavy-ion collisions. The authors have argued that a HIC environment is the ``most universal eraser" of kinematic effects. The high temperatures affect the masses and widths of the particles participating in the process and if a triangle reaction is completed during the fireball lifetime, the triangle singularities will be washed out in the hot hadron gas phase.  As a consequence, if the $Z_c$ is generated by a $D_1 D^* D^*$ triangle loop, it will disappear at temperatures close to the deconfinement temperature. Hence,  any future detection in heavy-ion collisions might be an indication that the $Z_c$ is indeed a real hadron, while its absence points towards the singularity interpretation.



\begin{acknowledgements}
 
This work was partly supported  by the Brazilian agencies CNPq/FAPERJ under the Project INCT-Física Nuclear e Aplicações (Contract No. 464898/2014-5). The work of L.M.A. is partly supported by the Brazilian agency CNPq under contracts 309950/2020-1, 400215/2022-5 and 200567/2022-5.


\end{acknowledgements} 




\begin{thebibliography}{99}

\bibitem{brambilla2019xyz}
N.~Brambilla, S.~Eidelman, C.~Hanhart, A.~Nefediev, C.-P. Shen, C.~E. Thomas, A.~Vairo, and C.-Z. Yuan, 
Phys. Rept. \textbf{873}, 1 (2020).

\bibitem{wang2022radiative}
X.-Y. Wang, G.~Li, C.-S. An, and J.-J. Xie, 
Phys. Rev. D {\bf 106},  074026 (2022).

\bibitem{ablikim2013observation}
M.~Ablikim, {\em et~al.}, 
Phys. Rev. Lett. {\bf 110}, 252001 (2013).

\bibitem{liu2013study}
Z.~Liu,  {\em et~al.},  Phys. Rev. Lett. {\bf 110}, 252002 (2013).

\bibitem{BESIII:2017bua}
M.~Ablikim {\em et~al.}, 
Phys. Rev. Lett.  {\bf 119}, 072001 (2017).

\bibitem{ablikim2015neutro}
M.~Ablikim {\em et~al.}, 
Phys. Rev. Lett. {\bf 115},  112003 (2015).

\bibitem{ablikim2015observation}
M.~Ablikim, {\em et~al.}, 
Phys. Rev. Lett.  {\bf 115}, 222002 (2015).

\bibitem{ablikim2014observation}
M.~Ablikim, {\em et~al.},          
Phys. Rev. Lett.  {\bf 112},  022001 (2014).

\bibitem{xiao2013observation}
T.~Xiao, S.~Dobbs, A.~Tomaradze, and K.~K. Seth, 
Phys. Lett. B \textbf{727}, 366 (2013). 


\bibitem{Workman:2022ynf}
R.~L. Workman {\em et~al.}, 
PTEP  {\bf 2022},  083C01 (2022). 

\bibitem{PhysRevD.88.054007}
F.-K. Guo, C.~Hidalgo-Duque, J.~Nieves, and M.~Pav\'on~Valderrama,
Phys. Rev. D  {\bf 88},  054007 (2013).

\bibitem{chen2016hidden}
H.-X. Chen, W.~Chen, X.~Liu, and S.-L. Zhu, 
Phys. Rept. {\bf 639},  1 (2016).

\bibitem{ChenAndDong}
D.-Y. Chen and Y.-B. Dong, 
Phys. Rev. D \textbf{93}, 014003 (2016). 

\bibitem{sun2011z}
Z.-F. Sun, J.~He, X.~Liu, Z.-G. Luo, and S.-L. Zhu,  Phys.  Rev.  D  {\bf 84}, 054002 (2011).

\bibitem{sun2012note}
Z.-F. Sun, Z.-G. Luo, J.~He, X.~Liu, and S.-L. Zhu,  Chin. Phys. C  {\bf 36},  194 (2012).

\bibitem{wang2014possible}
Z.-G. Wang and T.~Huang, Eur. Phys. J. C \textbf{74},  2891 (2014).

\bibitem{chen2015mass}
W.~Chen, T.~Steele, H.-X. Chen, and S.-L. Zhu,
Phys. Rev.  D {\bf 92}, 054002  (2015).

\bibitem{wilbring2013electromagnetic}
E.~Wilbring, H.-W. Hammer, and U.-G. Mei{\ss}ner, 
Phys. Lett. B  {\bf 726},   326  (2013).

\bibitem{dong2013strong}
Y.~Dong, A.~Faessler, T.~Gutsche, and V.~E. Lyubovitskij, Phys. Rev.  D {\bf 88}, 014030 (2013).


\bibitem{dong2014selected}
Y.~Dong, A.~Faessler, T.~Gutsche, and V.~E. Lyubovitskij, 
Phys. Rev.  D {\bf 89}, 034018 (2014).


\bibitem{li2014more}
G.~Li, X.~H. Liu, and Z.~Zhou, 
Phys. Rev.  D {\bf 90}, 054006 (2014).


\bibitem{gutsche2014radiative}
T.~Gutsche, M.~Kesenheimer, and V.~E. Lyubovitskij, 
Phys. Rev.  D {\bf 90}, 094013 (2013).


\bibitem{esposito2015probing}
A.~Guerrieri and A.~Pilloni, 
Phys. Lett. B  {\bf 746}, 194  (2015).

\bibitem{ke2013z}
H.-W. Ke, Z.-T. Wei, and X.-Q. Li, 
Eur. Phys. J. C \textbf{73},  2561 (2013).


\bibitem{Ji:2022uie}
T.~Ji, X.-K. Dong, M.~Albaladejo, M.-L. Du, F.-K. Guo, and J.~Nieves,
Phys. Rev.  D {\bf 106}, 094002 (2022).


\bibitem{Yan:2023bwt}
L.-W. Yan, Z.-H. Guo, F.-K. Guo, D.-L. Yao, and Z.-Y. Zhou, 
arXiv:2307.12283 [hep-ph].

\bibitem{maiani2005diquark}
L.~Maiani, F.~Piccinini, A.~Polosa, and V.~Riquer, 
Phys. Rev.  D {\bf 71}, 014028 (2005).

\bibitem{maiani2013j}
L.~Maiani, V.~Riquer, R.~Faccini, F.~Piccinini, A.~Pilloni, and A.~Polosa,
Phys. Rev.  D {\bf 87}, 111102 (2013).

\bibitem{Ali:2011ug}
A.~Ali, C.~Hambrock, and W.~Wang,  
Phys. Rev.  D {\bf 85}, 054011 (2012).

\bibitem{deng2014interpreting}
C.~Deng, J.~Ping, H.~Huang, and F.~Wang,  
Phys. Rev.  D {\bf 90}, 054009 (2014).
  
\bibitem{Wang2018}
Z.-G. Wang and J.-X. Zhang, Eur. Phys. J. C \textbf{78},  14 (2018).

\bibitem{dias2013z}
J.~M. Dias, F.~Navarra, M.~Nielsen, and C.~Zanetti, 
Phys. Rev.  D {\bf 88}, 016004 (2013).

\bibitem{Chen:2013coa}
D.-Y. Chen, X.~Liu, and T.~Matsuki, 
Phys. Rev.  D {\bf 88}, 036008 (2013).

\bibitem{Swanson:2014tra}
E.~S. Swanson, Phys. Rev.  D {\bf 91}, 034009 (2015).

\bibitem{Liu:2015taa}
X.-H. Liu, M.~Oka, and Q.~Zhao,  Phys. Lett. B  {\bf 753}, 297  (2016).  

\bibitem{Szczepaniak:2015eza}
A.~P. Szczepaniak,  Phys. Lett. B  {\bf 747}, 410  (2015).


\bibitem{Llanes-Estrada:2022qC}
Felipe J. Llanes-Estrada and Luciano M. Abreu, PANIC2021,  [arXiv:2110.14707 [hep-ph]]


\bibitem{CMS:2021znk}
A.~M.~Sirunyan \textit{et al.} [CMS], 
Phys. Rev. Lett. \textbf{128},  032001 (2022).


\bibitem{EXHIC}
S.~Cho {\em et~al.} [EXHIC], 
Prog. Part. Nucl. Phys. {\bf 95}, 279 (2017).


\bibitem{Abreu:2022lfy}
L.~M. Abreu, H.~P.~L. Vieira, and F.~S. Navarra, 
Phys. Rev. D {\bf 105},  116029  (2022).

\bibitem{abreu2018impact}
L.~Abreu, E.~Cavalcanti, and A.~Malbouisson,  
Nucl. Phys. A  {\bf 978}, 107  (2018); 
F.~S.~Navarra, M.~Nielsen, R.~S.~Marques de Carvalho and G.~Krein,
Phys. Lett. B {\bf 529}, 87 (2002); 
F.~O.~Duraes, S.~H.~Lee, F.~S.~Navarra and M.~Nielsen, 
Phys. Lett. B  {\bf 564}, 97 (2003). 


\bibitem{abreu2022exotic}
L.~M. Abreu, 
PoS  XVHadronPhysics, 012 (2022); arXiv:2201.07273 



\bibitem{abreu2022interactions}
L.~M. Abreu, F.~S. Navarra, M.~Nielsen, and H.~Vieira, 
Eur. Phys. J. C \textbf{82},  296 (2022). 

\bibitem{cho2013hadronic}
S.~Cho and S.~H. Lee, 
Phys. Rev.  C  {\bf 88}, 054901  (2013).

\bibitem{Nielson2007}
L.~W.~Chen, C.~M.~Ko, W.~Liu and M.~Nielsen,
Phys. Rev. C \textbf{76}, 014906 (2007). 


\bibitem{bracco2012charm}
M.~Bracco, M.~Chiapparini, F.~Navarra, and M.~Nielsen, 
Prog. Part. Nucl. Phys. \textbf{67}, 1019 (2012). 

\bibitem{oh2001j}
Y.~Oh, T.~Song, and S.~H. Lee, 
Phys. Rev. {\bf 63}, 034901 (2001).

\bibitem{carvalho2005hadronic}
F.~Carvalho, F.~Duraes, F.~Navarra, and M.~Nielsen,  
Phys. Rev. C {\bf 72}, 024902 (2005). 


\bibitem{ABREU2016303}
L.~Abreu, K.~Khemchandani, A.~M. Torres, F.~Navarra, and M.~Nielsen,
Phys. Lett. B {\bf 761}, 303  (2016).

\bibitem{Abreu1}
L.~M.~Abreu, F.~S.~Navarra, M.~Nielsen and H.~P.~L.~Vieira,
Phys. Rev. D \textbf{107}, 114013 (2023).


\bibitem{XProd1}        A. Martinez Torres, K. P. Khemchandani, F. S. Navarra, 
                        M. Nielsen and L. M. Abreu, 
                        Phys. Rev. D {\bf 90}, 114023  (2014).
                        

\bibitem{XProd2}        L. M. Abreu, K. P. Khemchandani, A. Martinez Torres, 
                        F. S. Navarra and M. Nielsen, 
                        Phys. Lett. B {\bf 761}, 303 (2016).

\bibitem{Abreu:2017cof}  L.~M.~Abreu, K.~P.~Khemchandani,
                         A.~Mart\'\i{}nez Torres, F.~S.~Navarra and M.~Nielsen,
                         Phys. Rev. C \textbf{97},  044902 (2018). 
                        

\bibitem{Abreu:2018mnc}  L.~M.~Abreu, F.~S.~Navarra and M.~Nielsen,
                         Phys. Rev. C \textbf{101},  014906 (2020).
                        


\bibitem{Cho:2015qca}
S.~Cho and S.~H.~Lee,
Phys. Rev. C \textbf{97}, 034908   (2018).


\bibitem{Abreu:2020ony}
L.~M.~Abreu,
Phys. Rev. D \textbf{103},   036013  (2021).


\bibitem{chiara} 
C.~Le Roux, F.~S.~Navarra and L.~M.~Abreu,
Phys. Lett. B \textbf{817}, 136284 (2021). 


\end{thebibliography}
\end{document}